\documentclass[conference]{IEEEtran}

\usepackage{amssymb}
\usepackage{amsmath}

\usepackage{fixmath}
\usepackage{graphicx}
\usepackage{psfrag}
\usepackage{ifthen}
\usepackage[amsmath,thmmarks]{ntheorem}
\usepackage{my_theorems}

\DeclareMathOperator{\Tr}{T}
\DeclareMathOperator{\tr}{tr}
\DeclareMathOperator{\He}{H}
\DeclareMathOperator{\Expect}{E}

\DeclareMathAlphabet{\mathbit}{OML}{cmr}{bx}{it}

\newcommand{\B}[1]{\mathbit{#1}}

\newcommand{\id}{\mathbf{I}}
\newcommand{\zero}{\mathbf{0}}

\newcommand{\Ptx}{P_{\mathrm{Tx}}}

\begin{document}

\title{An Asymptotic Analysis of the MIMO BC under Linear Filtering}

\author{\authorblockN{Raphael Hunger and Michael Joham}
\authorblockA{Associate Institute for Signal Processing, Technische
Universit{\"a}t
M{\"u}nchen,
80290 Munich, Germany\\
Telephone: +49 89 289-28508, Fax: +49 89 289-28504,
Email: \texttt{hunger@tum.de}}}

\maketitle

\begin{abstract}
  We investigate the MIMO broadcast channel in the high SNR regime
  when linear filtering is applied instead of dirty paper coding.
  Using a user-wise rate duality where the streams of 
  every single user are not treated as self-interference as
  in the hitherto existing stream-wise rate dualities for linear filtering,
  we solve the weighted
  sum rate maximization problem of the broadcast channel in the dual multiple
  access channel. Thus, we can exactly quantify the asymptotic rate loss of
  linear filtering compared to dirty paper coding for any channel realization. 
  Having converted the optimum covariance matrices 
  to the broadcast channel by means of the duality, we observe that
  the optimal covariance matrices
  in the broadcast channel feature quite complicated but still closed form 
  expressions although the respective
  transmit covariance matrices in the dual multiple access channel 
  share a very simple structure. We immediately come
  to the conclusion that block-diagonalization is the asymptotically
  optimum transmit strategy in the broadcast channel. Out of the set of
  block-diagonalizing precoders, we present the one which achieves the
  largest sum rate and thus corresponds to the optimum solution found
  in the dual multiple access channel.
  Additionally, we quantify the ergodic rate loss of linear coding compared
  to dirty paper coding for Gaussian channels with correlations at the mobiles.

\end{abstract}

\section{Introduction}

While the sum capacity of the single-user MIMO point-to-point link
can be expressed semi-analytically in closed form~\cite{Telatar}, 
the simplest multi-user setup with single antenna terminals already 
allows for the presumption
that this will remain infeasible in the broadcast and multiple access channel
irrespective of whether linear or nonlinear filtering is considered.
Fortunately, the high signal-to-noise ratio regime is an exception
to this deflating circumstance, since there, asymptotic results on the sum capacity
have been discovered for dirty paper coding and partly for linear
filtering.

\subsection{Literature Overview}

The single user point-to-point MIMO case was treated in~\cite{Salo_PtP,Salo_PtP_journal},
where the
\emph{Grant-Gauthier} lower bound on the mutual information, that becomes
asymptotically tight, was decomposed into 
a supremum capacity term, an instantaneous SNR effect term, and an
instantaneous capacity degradation term due to the eigenvalue spread.
Outage capacity and throughput of a fading point-to-point MIMO system are analyzed
in~\cite{Prasad_Isit}.
The first high-SNR sum capacity analysis of the point-to-multipoint broadcast channel appeared
in~\cite{Jindal_single_antennas}, where single-antenna receivers were considered.
Therein, the affine approximation of the sum capacity introduced in~\cite{Shamai_affine}
and elaborately discussed in~\cite{Lozano}
was utilized. First, \cite{Jindal_single_antennas} 
shows that the single-antenna broadcast channel has the same
asymptotic sum-capacity as the corresponding point-to-point MIMO link with cooperating
receive antennas, and second, how the power offset term in the broadcast channel looks like.
Furthermore, 
the instantaneous and ergodic spectral efficiency 
loss of linear zero-forcing beamforming with respect to DPC
was derived in~\cite{Jindal_single_antennas}, 
again for single antenna receivers.
The extension to multi-antenna receivers was presented in~\cite{DPCvsLin_Lee,Lee_MIMO},
where the asymptotic equivalence of the nonlinear dirty paper coding
sum capacity in the broadcast channel and
the sum capacity of the equivalent point-to-point MIMO link with cooperating receivers 
was proven to hold in the multi-antenna case.
Out of the class of linear precoding schemes, zero-forcing and block-diagonalization
are considered. However, only \emph{ergodic} statements for the asymptotic sum rate and
the asymptotic rate loss with respect to dirty paper coding are derived, and a very
special fading model is a key prerequisite for the presented results.
Expressions for the instantaneous rate loss are not possible. Moreover, it is neither
known yet, whether block-diagonalization is the asymptotically optimum transmission strategy
or not in the broadcast channel when linear filtering is considered, nor
how the optimum block-diagonalizing precoder looks like.

\subsection{Contributions}
The main contributions of this paper are summarized in the following list:
\begin{enumerate}
  \item The derivation of the maximum weighted sum rate asymptotically achievable with linear filtering.
  \item A closed form expression for the asymptotic rate loss of linear filtering 
		with respect to dirty paper coding for any antenna configuration at the base and
        the mobiles.
  \item A closed form solution of the covariance matrices in the dual uplink achieving this
		maximum weighted sum rate.
  \item We prove, that block diagonalization is asymptotically optimum in the broadcast channel.
  \item Finally, we derive the optimum precoding and transmit covariance matrices in the
        broadcast channel by means of our rate duality in~\cite{rate_duality_arxiv}.

\end{enumerate}


\section{System Model}
\label{sec:system_model}

We consider the communication between an $N$ antenna base station and
$K$ multi antenna terminals, where user~$k$ multiplexes~$B_k$ data streams
over his~$r_k$ antennas. For a short notation, we define $r$ as
the sum of all antennas at the terminals, i.e., $r=\sum_{k=1}^K r_k$,
and $b$ as the sum of all transmitted streams, i.e., $b=\sum_{k=1}^K B_k$.
In the MAC, user~$k$ applies a precoding matrix 
$\B{T}_k\in\mathbb{C}^{r_k\times B_k}$
generating his $r_k\times r_k$ transmit covariance matrix $\B{Q}_k=\B{T}_k\B{T}_k^{\He}$.
The precoded symbol vector propagates over the 
channel described by the matrix $\B{H}_k\in\mathbb{C}^{N\times r_k}$.
At the receiver side, zero-mean noise $\B{\eta}\in\mathbb{C}^N$ with identity
covariance matrix is added and the receive filter for user~$k$ is denoted 
by $\B{G}_k\in\mathbb{C}^{B_k\times N}$.
Due to the reversed signal flow in the BC, we characterize the transmission from the 
base station to terminal~$k$ by the Hermitian channel~$\B{H}_k^{\He}$
in the BC, the precoder dedicated to the $B_k$ streams of user~$k$
is denoted by $\B{P}_k\in\mathbb{C}^{N\times B_k}$, and zero-mean 
noise $\B{\eta}_k\in\mathbb{C}^{r_k}$ with identity covariance matrix
is added at user~$k$. Throughout this paper, we assume that the base station has at least as many antennas
as the terminals have in sum, i.e., $N\geq r$.

\section{Optimum Signalling in the Dual MAC}
\label{sec:OS_MAC}

Introducing the composite channel matrix $\B{H}$ and the
composite block-diagonal precoder matrix~$\B{T}$ of all $K$ users via
\begin{align*}
    \B{H} & = [\B{H}_1,\ldots,\B{H}_K] \in \mathbb{C}^{N\times r},
     \\
    \B{T} & =\boldsymbol{\operatorname{blockdiag}}\{\B{T}_k\}_{k=1}^K \in \mathbb{C}^{r\times b},
\end{align*}
the rate of user~$k$ seeing interference from all other users 
can be expressed as (see~\cite{rate_duality_arxiv})
\begin{equation}
  \begin{aligned}
      R_k & = \log_2\big|\id_{N}\!+\!\big(\id_N\!+\!\sum_{\ell\neq k}
		\B{H}_{\ell}\B{Q}_{\ell}\B{H}_{\ell}^{\He}\big)^{-1}
                       \!\B{H}_k\B{Q}_k\B{H}_k^{\He}\big| \\
     & = -\log_2\big|\id_{B_k} - \B{T}_k^{\He}\B{H}_k^{\He}\B{X}^{-1}\B{H}_k\B{T}_k\big|,
  \end{aligned}
  \label{rate_of_user}
\end{equation}
where the substitution $\B{X}$ reads as
\begin{equation*}
    \B{X}  =\id_N \!+\! \sum_{\ell=1}^K\B{H}_\ell\B{Q}_\ell\B{H}_{\ell}^{\He}
               = \id_N \!+\! \B{H}\B{T}\B{T}^{\He}\B{H}^{\He}.
\end{equation*}                
Reformulating the rate expression (\ref{rate_of_user}), we get 
\begin{equation}
  \begin{aligned}
    R_k & = -\log_2\big|\B{E}_k^{\Tr}\big(\id_b-\B{T}^{\He}\B{H}^{\He}\B{X}^{-1}
           \B{H}\B{T}\big)\B{E}_k\big| \\
           & = -\log_2\big|\B{E}_k^{\Tr}\big(\id_b+\B{T}^{\He}\B{H}^{\He}\B{H}\B{T}
               \big)^{-1}\B{E}_k\big|,
    \end{aligned}
\end{equation}
where the $k$th block unit matrix is defined via
\begin{equation*}
  \B{E}_k^{\Tr} = [\zero,\ldots,\zero,\id_{B_k},\zero,\ldots,\zero] \in \{0;1\}^{B_k\times b}
\end{equation*} 
with the identity matrix at the $k$th block.
Due to the assumption that the base station has more antennas than the terminals in sum,
all $r$ streams can be activated leading to square precoders $\B{T}_k$ with $B_k=r_k \ \forall k$.
Raising $\Ptx$, all $r$ streams become active, $\B{T}$ becomes full rank, and
all eigenvalues of $\B{T}^{\He}\B{H}^{\He}\B{H}\B{T}$ become much larger than one.
In the asymptotic limit, we obtain
\begin{equation}
  \begin{aligned}
    R_k & \cong -\log_2 \big|\B{T}_k^{-1}\B{E}_k^{\Tr}
        \big(\B{H}^{\He}\B{H}\big)^{-1}\B{E}_k\B{T}_k^{-\He} \big| \\
   & = \ \
     \log_2\big|\B{Q}_k\big| - \log_2 \big|\B{E}_k^{\Tr}
        \big(\B{H}^{\He}\B{H}\big)^{-1}\B{E}_k \big|,
  \end{aligned}
\end{equation}
since $\B{E}_k^{\Tr}\B{T}^{-1}=\B{T}_k^{-1}\B{E}_k^{\Tr}$.
The notation $x\cong y$ means that the difference $x-y$ vanishes when the sum power~$\Ptx$
goes to infinity.
Interestingly, the rate of user~$k$ depends only on the determinant of 
his own transmit covariance matrix $\B{Q}_k$, and not on the covariance
matrices of the other users! Consequently, the eigenbases of all transmit covariance matrices do not
influence the rates of the users, only the powers of the eigenmodes are relevant.
Let the eigenvalue decomposition of $\B{Q}_k$ read as $\B{Q}_k=\B{V}_k\B{\Lambda}_k\B{V}_k^{\He}$
with unitary $\B{V}_k$ and the diagonal nonnegative power allocation $\B{\Lambda}_k$.
Due to the determinant operator, $\B{V}_k$ can be chosen arbitrarily and therefore, 
we set $\B{V}_k=\id_{r_k} \ \forall k$ without loss of generality. Let the
power allocation matrix be composed by the entries
$\B{\Lambda}_k=\boldsymbol{\operatorname{diag}}\{\lambda_k^{(i)}\}_{i=1}^{r_k}$.
Due to the sum-power constraint, the determinant $|\B{Q}_k|=|\B{\Lambda}_k|$ is then maximized
by setting
\begin{equation}
\lambda_k^{(1)}=\ldots=\lambda_k^{(r_k)} := \lambda_k,
\end{equation}
i.e., by evenly distributing the power allocated to that user 
onto his individual modes, so 
$\B{Q}_k=\lambda_k \id_{r_k} \ \forall k$ with the sum-power constraint
  $\sum_{k=1}^Kr_k\lambda_k = \Ptx$.
Introducing nonnegative weight factors $w_1,\ldots,w_K$ for the rates
of the users, the weighted sum rate asymptotically reads as
\begin{equation}
    \sum_{k=1}^K\! w_k R_k\! \cong\! 
  \sum_{k=1}^K\! w_k\! \big(r_k \log_2\!\lambda_k 
  \!-\! 
  \log_2\! \big|\B{E}_k^{\Tr}
        \big(\B{H}^{\He}\B{H}\big)^{-1}\!\B{E}_k \big|\big).
  \label{asym_weighted}
\end{equation}
Subject to the sum power constraint $\sum_{k=1}^Kr_k\lambda_k = \Ptx$, the weighted sum rate in
(\ref{asym_weighted}) is maximized for 
\begin{equation}
  \lambda_k = \frac{w_k}{\sum_{\ell=1}^K w_\ell r_\ell}\Ptx,
\end{equation}
so the power is allocated to the users according to their weights
(similar to the single-antenna case proven in~\cite{DPCvsLin_Lee}),
and every user evenly distributes his fraction of power onto his modes.
In the case of identical weights $w_k=1 \ \forall k$,
the conventional sum rate asymptotically reads as
\begin{equation}
    \sum_{k=1}^K R_k \cong 
      r\log_2 \Ptx - r\log_2r
      -\sum_{k=1}^K \log_2 \big|\B{E}_k^{\Tr}
          \big(\B{H}^{\He}\B{H}\big)^{-1}\B{E}_k \big|,
  \label{asymptotic_rate_final}
\end{equation}
and is achieved with $\B{Q}_k=\Ptx/r\cdot\id_{r_k} \ \forall k$.
So, we are able to quantify the asymptotic sum rate
that can be achieved by means of linear filtering for every single channel realization
and antenna/user profile
in terms of the transmit power~$\Ptx$ and the channel itself.
In principle, the ergodic rate and the ergodic rate offset to dirty paper coding can
be obtained by averaging corresponding to \emph{any} distribution of the channel.
In~\cite{Lee_MIMO}, results on the \emph{ergodic} rate offset with respect to dirty paper coding
were presented for the specific case of Rayleigh fading only, 
where the channel entries of $\B{H}_1,\ldots,\B{H}_K$
all have the same distribution. Simple near-far effects with different average
channel powers for example cannot be captured
due to this restricting assumption. Moreover, the instantaneous rate offset expression is given
by means of bases representing null spaces of shortened channel matrices
taken from \cite{Spencer_ZF}
and not as a function of the channel purely as we do in (\ref{asymptotic_rate_final}).

Concerning the asymptotic rate expressions, we have now created a smooth 
transition from the $r$ single-antenna-users system configuration in~\cite{Jindal_single_antennas} where 
no cooperation exists between the antenna elements at the terminals, to the 
single-user point-to-point MIMO link where all $r$ antennas fully cooperate, see~\cite{Telatar} for example.
In between, we can now specify any antenna/user profile we want and compute the feasible rate
in the asymptotic limit. Using dirty paper coding, the asymptotic sum rate reads as
\begin{equation}
  \sum_{k=1}^K R_k^{\mathrm{DPC}} \cong r\log_2\Ptx - r\log_2 r + \log_2\big|\B{H}^{\He}\B{H}\big|
  \label{DPC_rate}
\end{equation}
and corresponds to the rate of the fully cooperating point-to-point link \cite{Lee_MIMO}.
Combining (\ref{DPC_rate}) and (\ref{asymptotic_rate_final}), the rate loss 
$\Delta R = \sum_{k=1}^K(R_k^{\mathrm{DPC}}-R_k)$ of 
optimal linear filtering with respect to optimal
dirty paper coding reads as
\begin{equation}
  \Delta R\cong \sum_{k=1}^K\log_2\!\big|\B{E}_k^{\Tr}(\B{H}^{\He}\B{H})^{-1}\B{E}_k\big|
  - \log_2\!\big|(\B{H}^{\He}\B{H})^{-1}\big|,
  \label{rate_difference}
\end{equation}
which of course vanishes, if all channels are pairwise orthogonal,
i.e., if $\B{H}^{\He}\B{H}$ is block-diagonal.
Of course, a block-type Hadamard inequality quickly leads to the inequality
\begin{equation*}
  - \log_2\big|(\B{H}^{\He}\B{H})^{-1}\big| \geq
  - \sum_{k=1}^K\log_2\big|\B{E}_k^{\Tr}(\B{H}^{\He}\B{H})^{-1}\B{E}_k\big|,
\end{equation*}
so linear filtering is obviously inferior to dirty paper coding.

\section{Optimum Signalling in the BC}
\label{sec:OS_BC}

Using our rate duality in \cite{rate_duality_arxiv}, we can convert the simple solution
for the covariance matrices in the dual MAC to covariance matrices in the BC, where the
Hermitian channels are applied. Since this duality explicitly uses the receive filters
in the MAC as scaled transmit matrices in the BC,
we first compute the MMSE receivers in the dual MAC, as they are optimum and generate
sufficient statistics. The receiver $\B{G}_k$ for user~$k$ in the dual MAC reads as
\begin{equation*}
 \B{G}_k = \B{E}_k^{\Tr}\B{T}^{\He}\B{H}^{\He}\big(\id_N+\B{H}\B{T}\B{T}^{\He}\B{H}^{\He}\big)^{-1}.
\end{equation*}
With the asymptotically optimum precoders $\B{T}_k\!=\!\sqrt{\Ptx/r}\id_{r_k}$, above expression asymptotically converges to
\begin{equation}
  \B{G}_k \cong \sqrt{r/\Ptx}\cdot\B{E}_k^{\Tr}\big(\B{H}^{\He}\B{H}\big)^{-1}\B{H}^{\He}.
  \label{asymptotic_G}
\end{equation}
Let $\B{P}_k$ denote the precoder of user~$k$ in the BC, then the $i$th column 
$\B{p}_{k,i}$ of $\B{P}_k$
follows from the conjugate $i$th row $\B{g}_{k,i}^{\prime\Tr}$ of the matrix
$\B{G}_k^\prime=\B{W}^{\He}_k\B{G}_k$ via (see~\cite{rate_duality_arxiv})
\begin{equation}
  \B{p}_{k,i} = \alpha_{k,i}\B{g}_{k,i}^{\prime*}
  = \frac{\alpha_{k,i}}{\sqrt{\Ptx/r}} \cdot\B{H}\big(\B{H}^{\He}\B{H}\big)^{-1}\B{E}_k
   \B{W}_k\B{e}_i,
  \label{pki_precoder}
\end{equation}
where the scaling factor $\alpha_{k,i}$ is obtained by the duality transformation
and $\B{W}_k$ is a unitary decorrelation matrix.
Since we convert only the asymptotically optimum transmit precoders and receive filters,
the duality transformation from the MAC to the BC 
in~\cite{rate_duality_arxiv} drastically simplifies and can even be computed
in closed form. In particular, the matrices $\B{M}_{a,b}$ 
in \cite[Eq.~(23)]{rate_duality_arxiv} vanish
for~$a\neq b$ yielding a diagonal matrix~$\B{M}$ and therefore, the scaling factors read as
\begin{equation}
  \alpha_{k,i} = \frac{\sqrt{\Ptx/r}}{\|\B{g}_{k,i}^{\prime}\|_2}.
  \label{scaling_factor}
\end{equation}
In combination with (\ref{pki_precoder}), the $i$th column of the precoder associated to user~$k$ reads as
\begin{equation*}
  \B{p}_{k,i} =\sqrt{\Ptx/r}\cdot \frac{ \B{H}\big(\B{H}^{\He}\B{H}\big)^{-1}\B{E}_k\B{W}_k\B{e}_i}
					{\big\| \B{H}\big(\B{H}^{\He}\B{H}\big)^{-1}\B{E}_k
   \B{W}_k\B{e}_i\big\|_2},
\end{equation*}
generating the precoder matrix
\begin{equation}
  \B{P}_k = \sqrt{\Ptx/r} \cdot
   \B{H}\big(\B{H}^{\He}\B{H}\big)^{-1}\B{E}_k\B{W}_k \B{D}_k^{-1},
  \label{BC_precoder}
\end{equation}
where the $i$th diagonal element of the diagonal matrix $\B{D}_k$ is
\begin{equation}
 [\B{D}_k]_{i,i} = \sqrt{\B{e}_i^{\Tr}\B{W}_k^{\He}\B{E}_k^{\Tr}\big(\B{H}^{\He}\B{H}\big)^{-1}
       \B{E}_k\B{W}_k\B{e}_i}.
  \label{D_matrix}
\end{equation}
We can immediately see, that the precoding filters in (\ref{BC_precoder}) lead to a block
diagonalization of the transmission, since $\B{H}_{\ell}^{\He}\B{P}_k=\zero$ holds for $k\neq \ell$.
Next, the decorrelation matrix $\B{W}_k$ which enables the duality is usually
chosen as the eigenbasis of $\B{G}_k\B{H}_k\B{T}_k\cong \id_{r_k}$, which asymptotically coincides with
the identity matrix due to~(\ref{asymptotic_G}). Since all eigenvalues are identical to one, the
decorrelation matrices $\B{W}_k$ are not given a priori, but can easily be computed such that the BC features the
same sum rate as the dual MAC. By means of (\ref{BC_precoder}) and the block diagonalization property of the precoders, we obtain for user $k$'s  receive signal
\begin{equation}
  \B{y}_k = \B{H}_k^{\He}\B{P}_k\B{s}_k+\B{\eta}_k = \sqrt{\Ptx/r} \cdot
   \B{W}_k \B{D}_k^{-1}\B{s}_k + \B{\eta}_k,
  \label{received_signal_k}
\end{equation}
where $\B{\eta}_k\in\mathbb{C}^{r_k}$ is the noise 
and $\B{s}_k$ the symbol vector
of user~$k$ both having an identity covariance matrix.
From (\ref{received_signal_k}), the rate of user~$k$ achieved in the BC reads as
\begin{equation*}
  R_k = \log_2\Big|\id_{r_k} + \Ptx/r\cdot \B{W}_k\B{D}_k^{-2}\B{W}_k^{\He}\Big|,
\end{equation*}
which asymptotically converges to
\begin{equation}
  R_k \cong r_k\log_2\Ptx - r_k\log_2 r - \log_2 |\B{D}_k^2|.
  \label{rate_BC}
\end{equation}
Above expression is maximized, if we choose $\B{W}_k$ as the unitary eigenbasis of
$\B{E}_k^{\Tr}(\B{H}^{\He}\B{H})^{-1}\B{E}_k$, see (\ref{D_matrix}), such that
$\B{D}_k^2$ contains the eigenvalues, i.e., the elements of $\B{D}_k^2$ are as different
as possible.
Thus, the transmit covariance matrix $\B{S}_k=\B{P}_k\B{P}_k^{\He}$ of user~$k$ reads as
\begin{equation}
  \B{S}_k = \frac{\Ptx}{r} \cdot \B{H}^{+\He}\B{E}_k
  \big(\B{E}_k^{\Tr}(\B{H}^{\He}\B{H})^{-1}\B{E}_k\big)^{-1}\B{E}_k^{\Tr}\B{H}^+
  \label{BC_covariance}
\end{equation}
with the channel pseudo-inverse $\B{H}^+=(\B{H}^{\He}\B{H})^{-1}\B{H}^{\He}$.
Note that $r_k$ eigenvalues of $\B{S}_k$ are $\Ptx/r$ whereas the remaining $N-r_k$ 
ones are zero. Thus, $\B{S}_k$ is a weighted orthogonal projector. 
Furthermore, $\tr(\B{S}_k)=\Ptx/r \ \forall k$, so the power is uniformly
allocated to the individual users in the broadcast channel as well.
Comparing (\ref{BC_covariance}) with the simple solution of the 
transmit covariance matrix $\B{Q}_k=\Ptx/r\cdot\id_{r_k}$ in the dual MAC,
it becomes obvious that the optimum covariance matrices 
are much more difficult to find directly in the BC without using the rate duality,
than in the dual MAC.
Plugging the optimum $\B{D}_k^2$ containing the eigenvalues of $\B{E}_k^{\Tr}(\B{H}^{\He}\B{H})^{-1}\B{E}_k$
into (\ref{rate_BC}) finally yields
\begin{equation}
  R_k \cong r_k\log_2\Ptx - r_k\log_2 r - \log_2
  \big|\B{E}_k^{\Tr}(\B{H}^{\He}\B{H})^{-1}\B{E}_k\big|.
  \nonumber
\end{equation}
Hence, the maximum sum rate (\ref{asymptotic_rate_final}) 
in the dual MAC is also achieved in the BC.

\section{Ergodic Rate Expressions}
\label{sec:ergodic}

In this section, we derive expressions for the asymptotic sum rate
when averaging over the channel realizations. The simple channel model
in~\cite{Lee_MIMO,Jindal_single_antennas} is a prerequisite for
the application of the ergodic analysis
due to the fact that an instantaneous analysis is not possible there.
We choose a more realistic channel where
near-far effects and channel correlations at the terminals are modeled as well,
i.e., the individual users can also have different average channel powers.
Thanks to our closed form expression of the
maximum asymptotic rate for an instantaneous channel realization,
the following ergodic analysis
is basically feasible for \emph{any} distribution of the channel
coefficients.
The channel matrices of the chosen near-far channel model 
with transmit correlations (in the MAC)
are defined by
$\B{H}_k = \bar{\B{H}}_k \B{C}_k^{\frac{1}{2}}\ \forall k$, where the elements of
$\bar{\B{H}}_k$ are uncorrelated and share a zero-mean i.i.d. Gaussian distribution
with variance one, and the Hermitian matrix $\B{C}_k^{\frac{1}{2}}$ contains
the correlations. An uncorrelated channel purely modeling the near-far effect
can be obtained by setting $\B{C}_k = c_k\id_{r_k}$, 
where $c_k>0$ is then the inverse path loss of user~$k$.
Let the $r\times r$ matrix $\B{C}$ be defined via
\begin{equation*}
  \B{C} = \boldsymbol{\operatorname{blockdiag}}\{\B{C}_k\}_{k=1}^K,
\end{equation*}
then the frequently arising inverse of $\B{H}^{\He}\B{H}$ reads as
\begin{equation*}
  (\B{H}^{\He}\B{H})^{-1} = \B{C}^{-\frac{1}{2}}(\bar{\B{H}}{}^{\He}\bar{\B{H}})^{-1}
  \B{C}^{-\frac{1}{2}},
\end{equation*}
where $\bar{\B{H}}^{\He}\bar{\B{H}}\sim \mathcal{W}_r(N,\id_r)$ has a \emph{Wishart}
distribution with $N$ degrees of freedom and 
$(\bar{\B{H}}{}^{\He}\bar{\B{H}})^{-1}\sim\mathcal{W}^{-1}_r(N,\id_r)$ has an \emph{inverse Wishart}
distribution, see~\cite{MV_Statistical,Matrix_Variate}. 
Thus, the ergodic value for the channel dependent log-summand
in the DPC sum rate expression~(\ref{DPC_rate}) reads as~\cite{Random_Matrix}
\begin{equation}
  \Expect\big[\log_2\big|\B{H}^{\He}\B{H}\big|\big] = 
  \frac{1}{\ln 2}\sum_{\ell=0}^{r-1}\psi(N-\ell)
  + \sum_{k=1}^K \log_2 |\B{C}_k|,
  \label{DPC_ergodic}
\end{equation}
where the \emph{Digamma}-function $\psi(\cdot)$ with integer arguments is defined 
via~\cite{Random_Matrix}
\begin{equation}
  \psi(n+1) = \psi(n) + \frac{1}{n} \ \ \text{if} \ n\in\mathbb{N},\ \ \psi(1) = -\gamma,
  \label{Digamma}
\end{equation}
and $\gamma$ is the \emph{Euler-Mascheroni} constant.
Note from~(\ref{DPC_ergodic}) that different path losses and correlations in the channel coefficients
simply lead to a shift of the
asymptotic rate curve.
Concerning the rate expressions with linear filtering, 
we exploit the property that the $k$th main diagonal block
of $(\bar{\B{H}}^{\He}\bar{\B{H}})^{-1}$ is also inverse
Wishart~\cite{Matrix_Variate}:
\begin{equation*}
  \B{E}_k^{\Tr}(\bar{\B{H}}^{\He}\bar{\B{H}})^{-1}\B{E}_k \sim
  \mathcal{W}^{-1}_{r_k}(N-r+r_k,\id_{r_k}),
\end{equation*}
In combination with 
$\B{E}_k^{\Tr}\B{C}^{-\frac{1}{2}}=\B{C}_k^{-\frac{1}{2}}\B{E}_k^{\Tr}$, this leads
to the ergodic expression
\begin{equation}
  \begin{aligned}
  & \Expect\big[ 
  \log_2\big|\B{C}_k^{-\frac{1}{2}}
  \B{E}_k^{\Tr}(\bar{\B{H}}^{\He}\bar{\B{H}})^{-1}\B{E}_k\B{C}_k^{-\frac{1}{2}}
  \big|
  \big] = \\ & \quad - \log_2 |\B{C}_k| 
   - \frac{1}{\ln 2}
   \sum_{\ell=0}^{r_k-1} \psi(N-r+r_k-\ell).
  \end{aligned}
  \label{linear_ergodic}
\end{equation}
By means of (\ref{DPC_ergodic}) and (\ref{linear_ergodic}), 
averaging over the asymptotic rate
difference $\Delta R$ in~(\ref{rate_difference}) between
linear filtering and DPC yields 
\begin{equation}
  \Expect[\Delta R] \cong
  \frac{1}{\ln 2}\Big[\sum_{\ell=0}^{r-1}\psi(N\!-\!\ell)
 -\! \sum_{k=1}^K\sum_{\ell=0}^{r_k-1} 
  \psi(N\!-\!r\!+\!r_k\!-\!\ell)\Big],
  \label{general_erg_loss}
\end{equation}
from which we can observe that the near-far effect with different path
losses and channel correlations does not influence the rate difference, since both
DPC and linear filtering are affected in the same way.

The general expression (\ref{general_erg_loss}) 
for the ergodic rate loss $\Expect[\Delta R]$
can be simplified by means of~(\ref{Digamma}), 
when all users are equipped with the same number of antennas. 
For the first special case, assume that each user has $\bar{r}>1$ antennas, i.e.,
$r_1=\ldots=r_K=\bar{r}$, such that the total number of antennas therefore
is $r=K\bar{r}$. After some manipulations, we obtain
\begin{equation}
  \Expect[\Delta R] \cong \frac{1}{\ln 2}
   \bigg[\sum_{\ell=1}^{(K-1)\bar{r}}\frac{\ell}{N-\ell}
    + \sum_{\ell=1}^{\bar{r}-1}\frac{(K-1)\ell}{N-K\bar{r}+\ell}   \bigg],
  \label{equal_ant_ergodic_rateloss}
\end{equation} 
which coincides with the results in~\cite{DPCvsLin_Lee,Lee_MIMO}, but is a different 
representation. For convenience, we assume that the summation vanishes if the upper limit of
a sum is smaller than the lower one, which happens for $\bar{r}=1$.
In this second special case with single antenna receivers,
i.e., $\bar{r}=1=r_k \ \forall k$ and $r=K$, the second sum
in~(\ref{equal_ant_ergodic_rateloss})
consequently vanishes, and the ergodic rate loss simplifies to
\begin{equation}
  \Expect[\Delta R] \cong \frac{1}{\ln 2}
  \sum_{\ell=1}^{K-1} \frac{\ell}{N-\ell},
  \label{single_ant_ergodic_rateloss}
\end{equation}
which is also a result of~\cite{Jindal_single_antennas}.

\section{Numerical Examples}
\label{sec:numerical}

In Table~\ref{table:erg_loss}, we present the ergodic rate loss of linear filtering
with respect to dirty paper coding
for different parameters $N$, $K$, $\bar{r}$, $r_1$, and $r_2$, where
we employed (\ref{equal_ant_ergodic_rateloss})
and (\ref{single_ant_ergodic_rateloss}) for 
the case $\bar{r}=r_1=\ldots=r_K$ (cf.\ \cite{Jindal_single_antennas,DPCvsLin_Lee,Lee_MIMO})
and (\ref{general_erg_loss}) for the case of different numbers of antennas
$r_1$, $r_2$.
It can be seen that a fully loaded single antenna system with $K=N$ and $\bar{r}=1$
has to face a significant rate reduction when switching from nonlinear to linear filtering.
Moreover, comparing the $K=2$ and $\bar{r}=3$ system with the one where $K=3$ and $\bar{r}=2$,
we observe that the rate loss in the first system is only 
$65$ percent
of the one in the second
system for $N=6$. We can infer that fewer terminals with many
antennas have to face smaller rate losses than many terminals with only few antennas.

\begin{figure}[t]
  \centering
  \psfrag{a}[c][c]{$10 \log_{10}\Ptx / \mathrm{dB}$}
  \psfrag{b}[c][c]{Sum Rate $\cdot \mathrm{Hz\ s/bits}$}
  \psfrag{c}[c][c]{$\Expect[\Delta R]$}

  \psfrag{a1}[tr][tr]{{\footnotesize Sum Capacity DPC}}
  \psfrag{a2}[r][r]{{\footnotesize Sum Rate Linear}}
  \psfrag{a3}[r][r]{{\footnotesize Affine Approx.\ DPC}}
  \psfrag{a4}[r][r]{{\footnotesize Affine Approx.\ Linear}}

  \psfrag{-10}{{\footnotesize -10}}
  \psfrag{-5}{{\footnotesize -5}}
  \psfrag{0}{{\footnotesize 0}}
  \psfrag{5}{{\footnotesize 5}}
  \psfrag{10}{{\footnotesize 10}}
  \psfrag{15}{{\footnotesize 15}}
  \psfrag{20}{{\footnotesize 20}}
  \psfrag{25}{{\footnotesize 25}}
  \psfrag{30}{{\footnotesize 30}}
  \psfrag{35}{{\footnotesize 35}}
  \psfrag{40}{{\footnotesize 40}}

  \includegraphics[width=3.4in]{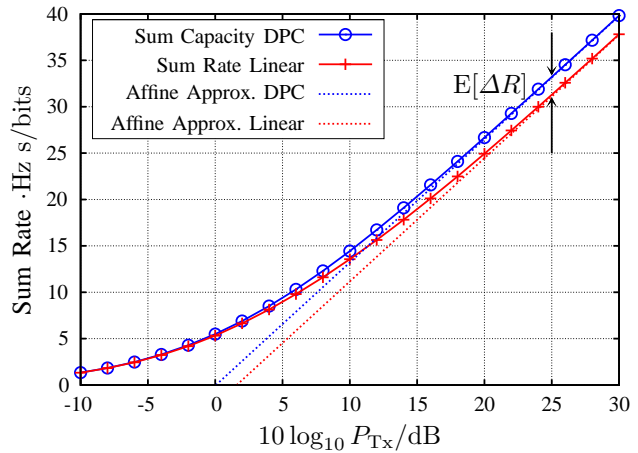}
  \caption{Ergodic sum rate of linear filtering and DPC and their respective affine approximations
           with $K=2$, $N=5$, and $r_1=r_2=\bar{r}=2$.}
  \label{fig:sum_rate_and_affine_approx}
\end{figure}

Next, we plot the ergodic sum capacity with DPC and the ergodic 
sum rate when linear filtering is
applied versus the transmit power~$\Ptx$ to see how large $\Ptx$ must be to let
the asymptotic affine approximations become tight. To this end, we choose a system
configuration where $K=2$ users each having $\bar{r}=2$ antennas are served by 
an $N=5$ antenna base station. Different path losses are modeled by setting 
$\B{C}_1=\id_2$ and $\B{C}_2=2\cdot\id_2$, i.e., user~$2$ has a stronger channel
on average, and we averaged over 1000 channel realizations.
While the DPC sum capacity can easily be computed via the algorithms
in~\cite{sum_power_iterative} or~\cite{HuScUt07}, an algorithm proven
to reach the maximum sum rate under
linear filtering does not seem to be available yet. Hence, we utilize
our combinatorial approach in~\cite{HuScJo08}, which obtains the 
best sum rate hitherto known in the case of linear filtering.
Fig.~\ref{fig:sum_rate_and_affine_approx}
shows that the asymptotic affine approximations become tight already for 
$\Ptx$ smaller than $20\mathrm{dB}$ and confirms the asymptotic ergodic rate loss
$\Expect[\Delta R]\cong 2.04$ from Table~\ref{table:erg_loss} which is
independent of the different average channel powers. 
For a multiplexing gain of $r=4$ as in the chosen system configuration,
this translates to an asymptotic power
loss of $1.54\mathrm{dB}$ of linear filtering with respect to DPC.

\section{Conclusion}
\label{sec:conclusion}

In this paper, we derived the asymptotic sum capacity which is maximally
achievable with \emph{linear} filtering in the broadcast channel by means of our
rate duality linking the rate region of the multiple access channel with the
broadcast channel rate region.
Due to the closed form expression of the asymptotic sum capacity for every single
channel realization, the instantaneous rate loss with respect to dirty paper
coding was presented, and the ergodic rate loss can quickly
be computed or simulated for any distribution of the fading process.
As an example, we presented the solution of the ergodic rate loss for
a simple fading model incorporating the near-far effect and correlations
at the mobiles.
Another key result proven is that block-diagonalization is the asymptotically
optimum transmission strategy in the broadcast channel. 
\begin{table}[t]
  \renewcommand\arraystretch{1.1}
  \tabcolsep7pt
  \centering
  \begin{tabular}{|c|ccccc|}
    \hline
     $K$,$\bar{r}$ & $N\!=\!2$ & $N\!=\!3$ & $N\!=\!4$ & $N\!=\!5$ & $N\!=\!6$ \\
    \hline
    \hline
    $2,1$ & $1.443$ & $0.721$ & $0.481$ & $0.361$ & $0.289$\\
    \hline
    $3,1$ & - & $3.607$ & $1.924$ & $1.322$ & $1.010$\\
    \hline
    $4,1$ & - & - & $6.252$ & $3.487$ & $2.453$\\
    \hline
    $5,1$ & - & - & - & $9.257$ & $5.338$\\
    \hline
    $6,1$ & - & - & - & - & $12.551$\\
    \hline
    $2,2$ & - & - & $3.366$ & $2.044$ & $1.491$\\
    \hline
    $2,3$ & - & - & - & - & $5.338$\\
    \hline
    $3,2$ & - & - & - & - & $8.223$\\
    \hline
    \hline
    $r_1,r_2$ & $N\!=\!2$ & $N\!=\!3$ & $N\!=\!4$ & $N\!=\!5$ & $N\!=\!6$\\
   \hline
   1,2 & - & $2.164$ & $1.202$ & $0.842$ & $0.649$\\
   \hline
   1,3 & - & - & $2.645$ & $1.563$ & $1.130$\\
   \hline
   1,4 & - & - & - & $3.006$ & $1.851$\\
   \hline
   2,3 & - & - & - & $4.208$ &$2.693$\\
   \hline
   2,4 & - & - & - & - & $4.857$\\
   \hline

  \end{tabular}
  \renewcommand\arraystretch{1.0}  
  \tabcolsep6pt
  \caption{Asymptotic ergodic rate loss $\Expect[\Delta R]$
           in $\mathrm{bits/s/Hz}$; $N$ antennas at the base, 
           $K$ users, $r_k$ antennas at user $k$.}
  \label{table:erg_loss}
\end{table}

\bibliographystyle{IEEEtran}
\bibliography{IEEEabrv,references}

\begin{thebibliography}{10}
\providecommand{\url}[1]{#1}
\csname url@rmstyle\endcsname
\providecommand{\newblock}{\relax}
\providecommand{\bibinfo}[2]{#2}
\providecommand\BIBentrySTDinterwordspacing{\spaceskip=0pt\relax}
\providecommand\BIBentryALTinterwordstretchfactor{4}
\providecommand\BIBentryALTinterwordspacing{\spaceskip=\fontdimen2\font plus
\BIBentryALTinterwordstretchfactor\fontdimen3\font minus
  \fontdimen4\font\relax}
\providecommand\BIBforeignlanguage[2]{{%
\expandafter\ifx\csname l@#1\endcsname\relax
\typeout{** WARNING: IEEEtran.bst: No hyphenation pattern has been}%
\typeout{** loaded for the language `#1'. Using the pattern for}%
\typeout{** the default language instead.}%
\else
\language=\csname l@#1\endcsname
\fi
#2}}

\bibitem{Telatar}
E.~Telatar, ``Capacity of multi-antenna gaussian channels,'' \emph{European
  Transactions on Telecommunications}, vol.~10, no.~6, pp. 585--596,
  November/December 1999.

\bibitem{Salo_PtP}
J.~Salo, P.~Suvikunnas, H.~M. El-Sallabi, and P.~Vainikainen, ``{Some results
  on MIMO mutual information: the high SNR case},'' in \emph{Global
  Telecommunications Conference (Globecom '04)}, vol.~2, December 2004, pp.
  943--947.

\bibitem{Salo_PtP_journal}
{J.~Salo and P.~Suvikunnas and H.~M.~El-Sallabi and P.~Vainikainen}, ``{Some
  Insights into MIMO Mutual Information: The High SNR Case},'' \emph{IEEE
  Transactions on Wireless Communications}, vol.~5, no.~11, pp. 2997--3001,
  November 2006.

\bibitem{Prasad_Isit}
N.~Prasad and M.~K. Varanasi, ``{Throughput analysis for MIMO systems in the
  high SNR regime},'' in \emph{International Symposium on Information Theory
  (ISIT)}, July 2006, pp. 1954--1958.

\bibitem{Jindal_single_antennas}
N.~Jindal, ``{High SNR Analysis of MIMO Broadcast Channels},'' in
  \emph{International Symposium on Information Theory (ISIT 2005)}, September
  2005, pp. 2310--2314.

\bibitem{Shamai_affine}
S.~Shamai and S.~Verd{\'u}, ``{The Impact of Frequency-Flat Fading on the
  Spectral Efficiency of CDMA},'' \emph{IEEE Transactions on Information
  Theory}, vol.~47, no.~4, pp. 1302--1327, May 2001.

\bibitem{Lozano}
A.~Lozano, A.~M. Tulino, and S.~Verd{\'u}, ``{High-SNR Power Offset in
  Multiantenna Communication},'' \emph{IEEE Transactions on Information
  Theory}, vol.~51, no.~12, pp. 4134--4151, December 2005.

\bibitem{DPCvsLin_Lee}
J.~Lee and N.~Jindal, ``{Dirty Paper Coding vs. Linear Precoding for MIMO
  Broadcast Channels},'' in \emph{40th Asilomar Conference on Signals, Systems,
  and Computers (Asilomar 2006)}, October 2006, pp. 779--783.

\bibitem{Lee_MIMO}
------, ``{High SNR Analysis for MIMO Broadcast Channels: Dirty Paper Coding
  Versus Linear Precoding},'' \emph{IEEE Transactions on Information Theory},
  vol.~53, no.~12, pp. 4787--4792, December 2007.

\bibitem{rate_duality_arxiv}
R.~Hunger and M.~Joham, ``{A General Rate Duality of the MIMO Multiple Access
  Channel and the MIMO Broadcast Channel},'' April 2008, \emph{Accepted for
  presentation at Globecom 2008}. Available at http://arxiv.org/abs/0803.2427.

\bibitem{Spencer_ZF}
Q.~H. Spencer, A.~L. Swindlehurst, and M.~Haardt, ``{Zero-Forcing Methods for
  Downlink Spatial Multiplexing in Multiuser MIMO Channels},'' \emph{IEEE
  Transactions on Signal Processing}, vol.~52, no.~2, pp. 461--471, February
  2004.

\bibitem{MV_Statistical}
R.~J. Muirhead, \emph{{Aspects of Multivariate Statistical Theory}},
  2nd~ed.\hskip 1em plus 0.5em minus 0.4em\relax Wiley, 2005.

\bibitem{Matrix_Variate}
A.~K. Gupta and D.~K. Nagar, \emph{{Matrix Variate Distributions}}.\hskip 1em
  plus 0.5em minus 0.4em\relax Chapman \& Hall /Crc, 1999.

\bibitem{Random_Matrix}
A.~M. Tulino and S.~Verdu, \emph{{Random Matrix Theory and Wireless
  Communications}}.\hskip 1em plus 0.5em minus 0.4em\relax Now Publishers Inc,
  2004.

\bibitem{sum_power_iterative}
N.~Jindal, W.~Rhee, S.~Vishwanath, S.~A. Jafar, and A.~J. Goldsmith, ``Sum
  {P}ower {I}terative {W}ater-{F}illing for {M}ulti-{A}ntenna {G}aussian
  {B}roadcast {C}hannels,'' \emph{{IEEE} Trans. Inform. Theory}, vol.~51,
  no.~4, pp. 1570--1580, 2005.

\bibitem{HuScUt07}
R.~Hunger, D.~A. Schmidt, and W.~Utschick, ``{Sum-Capacity and MMSE for the
  MIMO Broadcast Channel without Eigenvalue Decompositions},'' in \emph{IEEE
  International Symposium on Information Theory (ISIT)}, Nice, June 2007.

\bibitem{HuScJo08}
R.~Hunger, D.~A. Schmidt, and M.~Joham, ``{A Combinatorial Approach to
  Maximizing the Sum Rate in the MIMO BC with Linear Precoding},''
  \emph{Submitted to Asilomar 2008}.

\end{thebibliography}

\end{document}